# Starship Sails Propelled by Cost-Optimized Directed Energy

James Benford

Microwave Sciences

1041 Los Arabis Lane, Lafayette, CA 94549 USA

jbenford@earthlink.net

Microwave and laser-propelled sails are a new class of spacecraft using photon acceleration. It is the only method of interstellar flight that has no physics issues. Laboratory demonstrations of basic features of beam-driven propulsion, flight, stability ('beam-riding'), and induced spin, have been completed in the last decade, primarily in the microwave. It offers much lower cost probes after a substantial investment in the launcher. Engineering issues are being addressed by other applications: fusion (microwave, millimeter and laser sources) and astronomy (large aperture antennas). There are many candidate sail materials: carbon nanotubes and microtrusses, beryllium, graphene, etc. For acceleration of a sail, what is the cost-optimum high power system? Here the cost is used to constrain design parameters to estimate system power, aperture and elements of capital and operating cost. From general relations for cost-optimal transmitter aperture and power, system cost scales with kinetic energy and inversely with sail diameter and frequency. So optimal sails will be larger, lower in mass and driven by higher frequency beams. Estimated costs include economies of scale. We present several starship point concepts. Systems based on microwave, millimeter wave and laser technologies are of equal cost at today's costs. The frequency advantage of lasers is cancelled by the high cost of both the laser and the radiating optic. Cost of interstellar sailships is very high, driven by current costs for radiation source, antennas and especially electrical power. The high speeds necessary for fast interstellar missions make the operating cost exceed the capital cost. Such sailcraft will not be flown until the cost of electrical power in space is reduced orders of magnitude below current levels.

Laser and microwave propelled sails are a new class of spacecraft that uses photon acceleration. These *sailships* are the only method of interstellar flight that has *no physics issues*. Indeed, laboratory demonstrations of basic features of beam-driven propulsion have been completed in the last decade (primarily in the microwave). It offers much lower cost probes after a substantial investment in the launcher. *But how practical are its scale and economics?* This cost question has not been treated. So there is no understanding of what a lower-cost system would look like. nor have the variety of missions the beam-propelled sail method can embrace been explored from a cost point of view: Almost all studies deal with interstellar probes, the hardest of missions, but the far easier interplanetary missions have been little treated, even though they will come first. Only cost analysis can give concrete ideas of how to take the first steps in this new technology.

The cost is largely not the spacecraft, but the reusable launcher or 'beamer' (a system comprised of beam source and antenna(s) to radiate it). I derive general relations for cost-optimal transmitter aperture and beam power, then estimate both capital coat and operating cost of cost-transmitters using current cost parameters ($/W, $/m$^2$). Costs for large-scale manufacture of transmitters and antennas are well documented. (However, costs for space manufacture not known.) Below we account for economies of scale, which will be important, and characterize specific missions. In particular,

- Interstellar probes for exploration of the Oort Cloud, characterization of the nearby interstellar medium, and its interaction with the Heliosphere.
- Starships, either as primary propulsion of the mothership or as a means of decelerating probes from the mothership for Exoplanet exploration as the mothership flies on.

**1. Why Minimize Costs?**

Why consider economics at this early stage? Cost matters because it makes a big constraint, a game-changing difference. That's how we decide on competing claims for resources. Other starship propulsion methods, are typically nuclear: fission, fusion, their hybrid, and matter/anti-matter. Their costs are unquantifiable, although we can be sure each ship will be very expensive.

We ask: for a given kinetic in the launched sail, what is the cost-minimum beaming system? The usual method of cost optimization is to examine many alternative approaches to building a system, estimating the cost of each, and then comparing them. This is a 'bottom-up' approach. We offer a more general 'top-down' method based on analysis and actual experiences of designers. This gives a broader approach, capable of embracing new mission ideas as they arise. The cost of first steps in this field will be essential in planning how to proceed.

Cost of beaming systems is driven by two elements

- Capital cost $C_C$, divided into the cost of building the microwave source and the cost of building the radiating aperture $C_A$, and the
- Operating cost $C_O$, meaning the operational labor cost and the cost of the electricity to drive the system:

$$\begin{aligned} C &= C_C + C_O \\ C_C &= C_A + C_S \end{aligned} \quad (1)$$

Note that we'll find that, for high velocity missions, sails will have much smaller costs than the system that accelerates them. Consequently, sail cost is not included in the capital cost.

To optimize, meaning minimize, the cost, the simplest approach is to assume power-law scaling dependence on the peak power and antenna area. Using this cost-minimization approach, we have been able to correctly estimate (to within 15%) the actual cost of ORION, a transportable, self-contained HPM test facility first fielded in 1995 and currently in operation [1]

Examples of how cost optimization can make a big difference appear in Appendix A.

## 2 State of Knowledge

Microwave and Laser propelled sails are a new class of spacecraft that promises to revolutionize future space travel. (For a general introduction to solar and beam-propelled sails, the reader is referred to McInnes [2].)

### 2.1 Theory

The beam (microwave and laser)-driven sail spacecraft was first proposed by Robert Forward [3,4]. The sail acceleration $a_S$ from *photon momentum* produced by a power P on a thin sail of mass m and area A is

$$a_S = [\eta+1] \, P/\sigma A c \qquad (2)$$

where $\eta$ is the reflectivity of the film of absorbtivity $\alpha$, c the speed of light, $\sigma$ the area mass density, $m = \sigma A$, with A the sail area. . Note that we've neglected the payload mass by counting only the sail. This may not be true for smaller sails such as the interstellar precursor discussed in section 4.

The force from photon acceleration is weak, but is observed in the trajectory changes in interplanetary spacecraft, because the solar photons act on the craft for years. For solar photons, with power density ~1 kW/m$^2$ at Earth's orbit, current solar sail construction gives accelerations ~1 mm/s$^2$, a very low acceleration. Shortening mission time means using much higher power densities. In this case, to accelerate at one earth gravity requires ~10 MW/m$^2$. However, for launch from orbit into interplanetary and interstellar space, much lower accelerations, and hence much lower power densities are needed.

Of the power incident on the sail, a fraction $\alpha P$ will be absorbed. In steady state, this must be radiated away from both sides of the film, with an average temperature T, by the Stefan-Boltzmann law

$$\alpha P = 2 A \, \varepsilon \, \kappa \, T^4 \qquad (3)$$

where $\kappa$ is the Stefan-Boltzmann constant and $\varepsilon$ is emissivity. Eliminating P and A, the sail acceleration is

$$a_S = 2 \, \kappa/c \, [\varepsilon \, (\eta+1) \, /\alpha] \, (T^4/m) \qquad (4)$$

where we have grouped constants and material radiative properties separately. Clearly, the acceleration is strongly temperature-limited, $\sim T^4$. This fact means that materials with low melt temperatures (Al, Be, Nb, etc.) cannot be used for fast beam-driven missions. For example, aluminum has a limiting acceleration of 0.36 m/s$^2$, which is <4% of a gee. The invention of strong and light carbon mesh materials has made laboratory sail flight possible because carbon has no liquid phase, and sublimes instead of melting. Carbon can operate at very high temperature, up to 3000 C, and graphene paper could be well above 4000 K. Its limiting acceleration is in the range of 10-100 m/s$^2$, or 1-10 gees, sufficient to launch in vacuum (to avoid burning) in earth-bound laboratories.

## 2.2 Experiment

Recently, beam-driven sail flights have demonstrated the basic features of the beam-driven propulsion. This work was enabled by invention of strong, light carbon microtruss material, which operates at high temperatures to allow liftoff under one earth gravity. Experiments with carbon-carbon microtruss material driven by microwave and laser beams have observed flight of ultralight sails of at several gee acceleration [5]. In the microwave experiments, propulsion was from a 10 kW, 7 GHz beam onto sails of mass density 5g/m$^2$ in 10$^{-6}$ Torr vacuum. At microwave power densities of ~kW/cm$^2$, accelerations of several gravities were observed (Fig. 1). Sails so accelerated reached temperatures ~2000 K from microwave absorption, and remained intact. (Much lower power densities and accelerations are needed in the missions we'll analyze. We had to hit it powerfully because we needed >1 gee to lift off.)

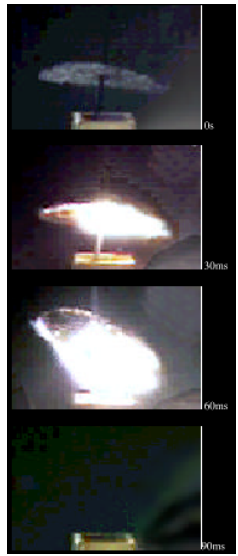

**Fig. 1.** Carbon sail lifting off of rectangular waveguide under 10 kW microwave power at 2 gees (four frames, first at top) in vacuum chamber [5]. Sail heats up, lifts off, and in the bottom frame the sail has flown away.

Will sails riding beams be stable? The requirement of *beam-riding*, stable flight of a sail propelled by a beam, places considerable demands on the sail shape. Even for a steady beam, the sail can wander off if its shape becomes deformed or if it does not have enough spin to keep its angular momentum aligned with the beam direction in the face of perturbations. Beam pressure will keep a concave shape sail in tension, and it will resist sidewise motion if the beam moves off-center, as a sidewise restoring force restores it to its position [6]. Experiments have verified that beam-riding does occur [7]. Positive feedback stabilization seems effective when the sidewise gradient scale of the beam is the same as the sail concave slope. A broad conical sail shape appears to work best.

The beam can also carry angular momentum and communicate it to a sail to help control it in flight. Circularly polarized electromagnetic fields carry both energy and angular momentum, which acts to produce a torque through an effective moment arm of a wavelength, so longer wavelengths are more efficient in producing spin. This effect can be used to stabilize the sail against the drift and yaw, which can cause loss of beam-riding, and allows 'hands-off' control of the sail spin, and hence stability, at a distance.

**2.3 Missions**

It's important to realize that for large-scale space power beaming to become a reality it must be broadly attractive. This means that it must provide for a real need, make business sense, attract investment, be environmentally benign, be economically attractive and have major energy or aerospace firms support and lobby for it. Therefore, we include missions that could lead to Starwisp missions, from an infrastructure base developed for smaller-scale missions.

<u>Interplanetary Launch</u> An early mission for microwave space propulsion is dramatically shortening the time needed for solar sails to escape Earth orbit. By sunlight alone, sails take about a year to climb out of the earth's gravity well. Computations show that a ground-based or orbiting transmitter can impart energy to a sail if they have resonant paths – that is, the beamer and sail come near each other (either with the sail overhead an Earth-based transmitter or the sail nearby orbits in space) after a certain number of orbital periods. For resonance to occur relatively quickly, specific energies must be given to the sail at each boost. If the sail is coated with a substance that sublimes under irradiation, much higher momentum transfers are possible, leading to further reductions in sail escape time. This new method, already shown in the laboratory, promises to greatly improve times to lift sails into interplanetary orbits. Simulations of sail trajectories and escape time are shown in Fig. 2. In general, resonance methods can reduce escape times from Earth orbit by over two orders of magnitude versus using sunlight alone on the sail [8].

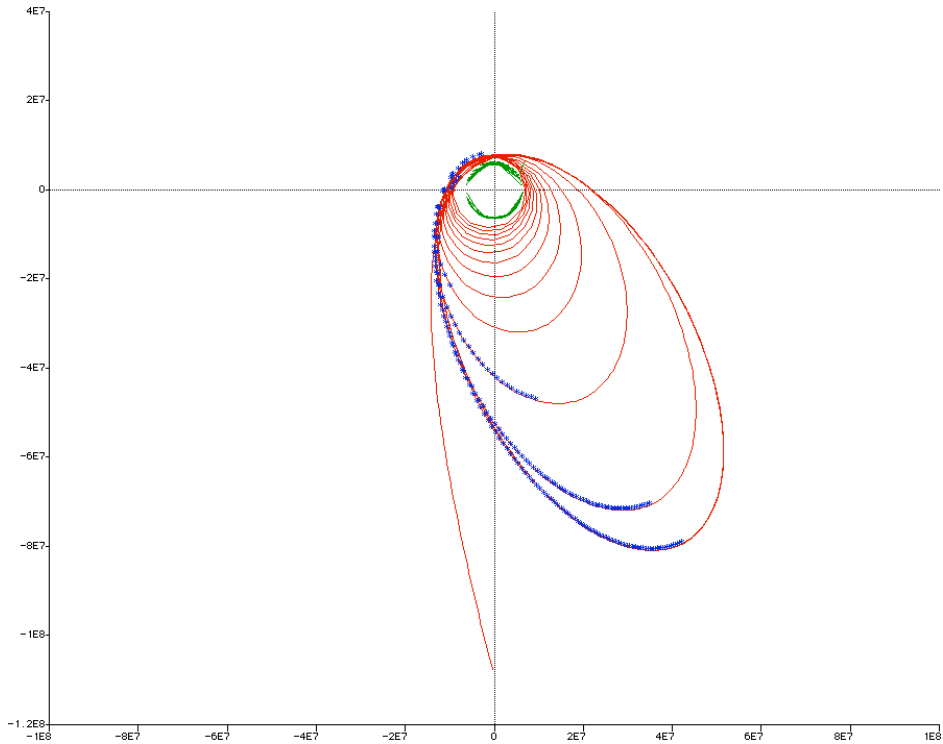

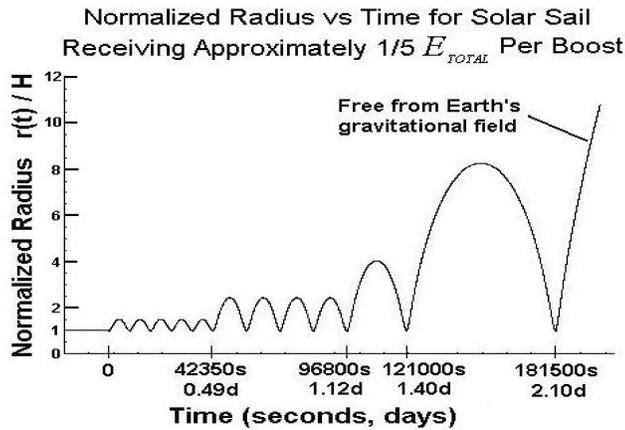

**Fig. 2** Beam-driven sail trajectory out of earth orbit. a) Simulation: time of beam acceleration is the thicker line. 2 b) Radius normalized to earth radius vs. time for sail [8].

While microwave transmitters have the advantage that they have been under development much longer than lasers and are currently much more efficient and inexpensive to build, they have the disadvantage of requiring much larger apertures for the same focusing distance. This is a significant disadvantage in missions that require long acceleration times with correspondingly high velocities. However, it can be compensated for with higher acceleration. The ability to operate carbon sails at high temperature enables much higher acceleration, producing large velocities in short distances, thus reducing aperture size. Very low-mass probes could be launched from Earth-based microwave transmitters with maximum acceleration achieved over a few hours using apertures only a few hundred meters across.

A number of such missions have been quantified. These missions are for high velocity mapping of the outer solar system, Jupiter, Kuiper Belt, Plutinos, Pluto and the Heliopause, and the interstellar medium. Meyer and co-workers [9] described an attractive interplanetary mission: *rapid delivery of critical payloads within the solar system*. For example, such emergencies as crucial equipment failures and disease outbreaks, can make fast delivery of small mass payloads to, say, Mars colonies, urgent. They describe missions with 175 km/sec speeds driven by lasers or microwaves at fast boost for a few hours of acceleration, coast at high speed, decelerate for a few hours into Mars orbit (by aerocapture or a decelerating beamer) -- transit time 10 days. The Benfords' *Mars Fast Track* then extended this to missions with 5 gee acceleration near Earth. Using a ground station, acceleration occurs for a couple of hours for a 100 kg payload. Jordin Kare quantified a Jupiter mission with beamed energy [10].

Interstellar Probes are solar/interstellar boundary missions out to ~1000 AU. The penultimate is the interstellar precursor mission. For this mission class, operating at high acceleration the sail size can be reduced to less than 100 m and accelerating power ~100 MW focused on the sail [11]. At 1GW, sail size extends to 200 m and super-light probes reach velocities of 250 km/s for very fast missions. In a NIAC study, McNutt and co-workers have described such missions driven by rocket and gravity assists. Beaming power could make for shorter mission durations. Here transit time is a serious factor driving mission cost.

Starships Truly the biggest and grandest mission. This concept requires very large transmitter antenna/lens and receiver (sail) optics (e.g., 1,000-km diameters for missions to 40 ly). A Space Solar Power station radiates a microwave beam to a perforated sail made of carbon nanotubes with lattice scale less than the microwave wavelength. The scale of the first concept was enormous, but Geoff Landis found ways to reduce it dramatically [12-15]. Systems much smaller than those of Forward were described by Frisbee [16], with peak power ~10 GW and size (~1km sail, ~1000 km antenna array aperture). Presumably, cost is also lowered, but has not been quantified. We describe here an economic approach to sailships for further reduction in power, size and cost.

**3 Cost-Optimization of Beam-Driven Sails**

**3.1 Capital Cost**

Whatever the mission, to optimize sailships, meaning to minimize the cost, the simplest approach is to assume power-law scaling dependence on the peak power P and antenna area A. This is a well-established method in industry [17]. The basic equation follows from linear scaling with coefficients describing the dependence of cost on area and radiated electromagnetic power. Antenna (or optic, in the case of lasers) areal cost coefficient a($/m$^2$) includes cost of the antenna, its supports and sub-systems for pointing and tracking and phase control, and is not to be confused with the sail acceleration $a_S$. Radiated power cost coefficient p ($/W) includes the source, power supply, cooling equipment and prime power cost.

$$C_A = aA$$
$$C_S = pP \quad (5)$$
$$C_C = aA + pP$$

I neglect any fixed costs, which would vanish when we differentiate to find the cost optimum. Mass production can decrease the cost of antenna elements and power modules, and will later be included in 'learning curve' factors in the coefficients a and p. We assign from mission requirements the beam frequency, final velocity, sail mass and diameter as an input parameter with respect to optimization.

**3.2 Capital Cost-Optimum Scaling**

I formulate the cost equation using equation A4 (in Appendix B) in terms of P, then substitute for $R_0$. Note that $V_0$ is the speed at $R_0$, the point where beam size exceeds sail size and the beam is switched off. This final speed, which when

$$P = \frac{V_0^2 mc}{R_0[\eta+1]} = \frac{2.44 V_0^2 mc^2}{2D_t D_s f[\eta+1]} \quad (6)$$

$$C_C = aA + pP = \frac{a\pi D_t^2}{4} + p\frac{2.44 V_0^2 mc^2}{2D_t D_s f[\eta+1]} \quad (7)$$

Collect constants into g:

$$C_C = \frac{a\pi D_t^2}{4} + gp\frac{V_0^2 mc^2}{D_t D_s f} \quad (8)$$

where

$$g = \frac{2.44}{2[\eta+1]} = \frac{1.22}{\eta+1} \cong 0.65 \quad (9)$$

for reflectivity slightly <1.

Differentiate with respect to transmitter diameter and set it equal to zero, giving the optimum diameter, power and area:

$$\frac{\partial C_C}{\partial D_t} = \frac{a\pi}{2} D_t - g\frac{pmV_0^2 c^2}{D_t^2 D_s f} = 0 \qquad (10)$$

$$D_t^{opt} = \left[\frac{2g}{\pi}\frac{p}{a}\frac{mV_0^2 c^2}{D_s f}\right]^{1/3} \qquad (11)$$

After some algebra,

$$C^{opt} = \frac{4}{3}\sqrt{\frac{5}{2}}\left(p^{2/3}a^{1/3}\right)\left[\frac{mV_0^2}{D_s f}\right]^{2/3}\frac{c^{4/3}}{(\eta+1)^{2/3}} \qquad (12)$$

$$C_P^{opt} = C_A^{opt}$$

*Minimum capital cost is achieved when the cost is equally divided between antenna gain and radiated power.* This derived result was used as a rule-of-thumb by microwave system designers for rough estimates of system cost. This cost ratio was independently discovered from cost data on the Deep Space Network [18]. For a recent example, Kare and Parkin have built a detailed cost model for a microwave beaming system for a beam-driven thermal rocket and compared it to a laser-driven rocket. They also find that, at minimum, cost is equally divided between the two cost elements [19]. However, the power beaming relation in reference 17 for $C^{opt}$ is different from Eq. 12. That's because here we add the sails acceleration dynamics, making the relation scale differently.

The optimum antenna area is

$$A^{opt} = \frac{C^{opt}}{2a} = \frac{2}{3}\sqrt{\frac{5}{2}}\left(\frac{p}{a}\right)^{2/3}\left[\frac{mV_0^2}{D_s f}\right]^{2/3}\frac{c^{4/3}}{(\eta+1)^{2/3}} \qquad (13)$$

Similarly, the optimum power is

$$P^{opt} = \frac{C^{opt}}{2p} \qquad (14)$$

And the relation between radiator power and the radiating aperture is:

$$A^{opt} = \frac{p}{a} P^{opt} \tag{15}$$

### 3.3 Operating Cost

The other cost element is the *operating cost $C_O$*. That is the cost of electricity to drive the beam sources, with a cost coefficient $p_{ave}$ ($/W), which at present in the US is 0.1 $/kW-hr = 2.8 $10^{-8}$ $/J or 36 MJ/$. There is some inefficiency in generating beam power, including voltage multiplication and source inefficiencies. In fully developed microwave system at 1-10 GHz, efficiency is $\varepsilon_b$ ~0.8, for millimeter wave beams at ~100 GHz, $\varepsilon_b$ ~0.4, but for lasers, can be $\varepsilon_b$ <0.1.

From Eq. 2,

$$m \int dv = mv_0 = \frac{[1+\eta]}{c} \int P dt = \frac{[1+\eta]}{c} E_b \tag{16}$$

where $E_b$ is the energy beamed at the sail. The efficiency $\varepsilon$ of power beaming is low because only momentum is transferred. The efficiency of producing kinetic energy KE ($= mv^2/2$) and total beamed energy required is,

$$\varepsilon = \frac{KE}{E_b} = \frac{1+\eta}{2}\frac{v}{c} \cong \frac{v}{c}$$
$$\frac{E_b}{KE} \cong \frac{c}{v} \tag{17}$$

The operating cost is then the product of kinetic energy and the cost of electrical energy divided by the efficiency:

$$C_O = p_{ave} \frac{E_b}{\varepsilon_b} = \frac{p_{ave}}{\varepsilon_b} \frac{c}{v}[KE] = \frac{p_{ave}}{\varepsilon_b} \frac{c}{v}\left[\frac{mv^2}{2}\right] \tag{18}$$

Electrical operating cost increases linearly with both speed and mass. For example, the KE of a kilogram at 0.1 c is 4.5 $10^{14}$ J, and for microwave beaming, the above i.e., the electrical cost to launch it is 158 M$ (assuming $p_{ave}$= 0.1 $/kW-hr). As shown in 4.3, speeds necessary for fast interstellar missions make the operating cost exceed the capital cost. A clear conclusion is that such missions will not be flown until the cost of electrical power in space is reduced orders of magnitude below current levels.

## 3.4 Economies of Scale, Learning Curve

The components we're modeling here, antennas and sources of microwave, mm-wave and laser beams, will be produced in large quantities for the large scales of directed energy-driven sails. High-volume manufacturing will drive costs down. Such economies of scale are accounted for by the *learning curve*, the decrease in unit cost of hardware with increasing production, shown in Fig. 3. This is expressed as the cost reduction for each doubling of the number of units, the learning curve factor $f$. This factor typically varies with differing fractions of labor and automation, $0.7 < f < 1$, the later value being total automation. A combination of both automation and hands-on fabrication gives 85%, which fits the data from power beaming technologies. The cost of N units is

$$C_N = C_1 N^{1+\frac{\log f}{\log 2}}$$

$$CSR = \frac{NC_1}{C_1 N^{1+\frac{\log f}{\log 2}}} = N^{-\frac{\log f}{\log 2}} \tag{19}$$

where $C_1$ is the cost of a single unit and the *cost savings ratio* CSR is the improvement gained by economies of scale, which depends only on the learning curve factor $f$ and the number of units.

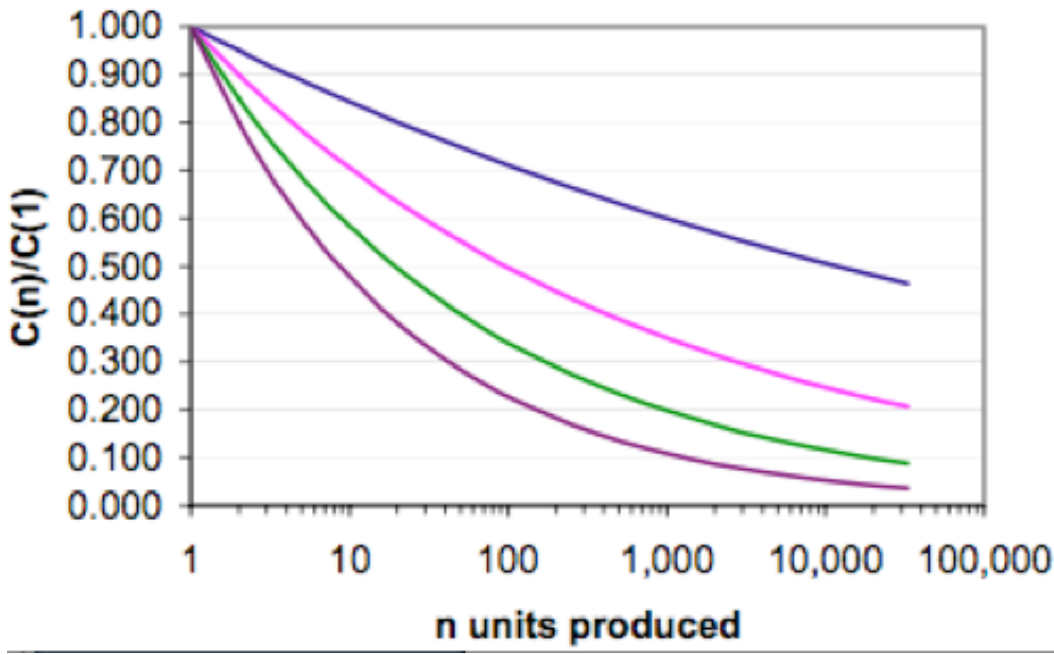

**Fig. 3.** Learning curves for four learning curve $f$'s: 95%, 90%, 85%. 80%. Power beaming technologies data fit the 85% curve.

For example, for a 90% learning curve, the cost of a second item is 90% of the cost of the first, the fourth is 90% of the cost of the second, and the $2N^{th}$ item is 90% of the cost of the $N^{th}$ item. Then the cost of 64 items is 64 times 0.53 $C_1$. Ordered in bulk, the 64 items will cost 47% less than ordered for production one-at-a-time. The CSR is 1.88. Thus, economies of scale reduce cost by larger and larger amounts as systems grow.

For the technologies of power beaming systems, it is well documented that *antennas and microwave sources have a 85% learning curve*, (learning curve factor $f$=0.85) based on large-scale production of antennas, magnetrons, klystrons and TWTs [21]. For from the above CSR=$N^{0.233}$, $C_N$ = $C_1 N^{1-0.233}$ = $C_1 N^{0.767}$.

Note that, because the number of elements of antennas and millimeter wave sources is different, the economies of scale are different, making their equal costs in the basic case (Eq. 12) substantially different when economies of scale are applied to bring down system cost (see Table 2).

## 4. Starship Examples

We work a few example missions, using the methodology developed here, to illustrate how the key features of Starships driven by beamed energy can be deduced from a few requirements in a self-consistent way, resulting in the lowest-cost concept. One then learns what factors drive a specific result. Then the assumed parameters, such as sail diameter, can be varied to get a more attractive system. See Appendix C for the methodology.

We consider only flyby probe missions, so no deceleration occurs. The key features are shown in Table 1. The much lower economies-of-scale costs, with learning curve 0.85 taken into account, are in Table 2, based on coefficients in Table 3.

Note one of the key challenges of directed energy is the pointing accuracy; i.e., that required to keep the beam on the sail. That parameter is roughly the ratio of the sail diameter to the accelerating distance, $D_S/R_0$.

### 4.1 Interstellar Precursor—$2 \times 10^{-4}$c—Getting our Feet Wet

What is the cost of a power beaming facility to lunch a probe out of the Solar System? Velocity is assumed to be 63 km/sec, as in studies of precursors [21]. We use the cost coefficients of present day millimeter wave gyrotron systems, p=3\$/W, and antennas for ~100 GHz, a=10,000 \$/m$^2$. For the sail characteristics, choose the sail mass/area $\sigma_S$ = 4 x10$^{-5}$ kg/m$^2$, and reflectivity $\eta$= 0.9, which are at the optimistic end of the values Matloff provides for a beryllium hollow-body sail [22]. The 1-km diameter sail mass is 31.7 kg. Assume the sail mass equals $m_P$ the payload mass, so m=2 $m_P$, total mass is 62.8 kg. Kinetic energy is roughly $10^{11}$ J. The beaming system has these derived parameters: Power 38 GW, antenna diameter 3.8 km, optimum capital cost 230 B\$ (without economies of scale, which reduce costs by one or two orders of magnitude, see below). The distance over which the sail is accelerated efficiently (Eq. A1) $R_0$ =520,000 km, f the order of the distance to the moon. Acceleration is 2.4 m/s$^2$, a quarter of a gee, and acceleration time is 5 hours. Pointing accuracy is µradians, present-day capability. So this mission can be done from a ground array, with power from the Earth grid, from a site with altitude high enough and humidity low enough that 100 GHZ mm-waves propagate through the atmosphere with little loss.

To gain economies-of-scale, gyrotron cost scales as the power P, and 1.2 MW units now cost 3.6 M$, or $3/W, including the power supply, in single units. The number of such units needed for 24 GW is 32,000. From Eq. 19, cost of the millimeter wave power will be reduced by economies of scale to 10 B$. Antennas in the 100 GHz range cost about $10,000/m$^2$, based on the 64 12-m diameter dishes of ALMA. The number of ALMA-sized elements needed is 1,600. The antenna cost falls to 19.5 B$. The total capital cost falls from 230 B$ to 30 B$, so cost savings ratio CSR~8. This is about ten times the cost of the Flagship missions to the outer planets, Galileo and Cassini.

The operating cost, i.e., the electrical cost to launch out of the Solar System is 41 M$. Once built, the beaming facility can send many probes into the interstellar medium. The strategy will be to use the system to launch sequences of sails in many directions to sample the Interstellar Medium and flyby Kuiper Belt and Oort Cloud objects, such as Senda. As the facility grows, the sails will be driven faster and can carry larger payloads.

## 4.2 Starship at 0.1c—Into the Interstellar

This Starship makes the big leap into the Interstellar with speed ~500 times that of the Precursor. Kinetic energy is 1.8 x 10$^{18}$ J. Pointing accuracy becomes tens of nanoradians, beyond present capability. The acceleration distance is four times the Sun-Earth distance, because of the low sail diameter. Consequently, the acceleration required is very high, about 80 gees. But from Eq. 4, acceleration is strongly temperature-limited. Not even carbon can survive the heating due to absorption at this acceleration. Sail reflectivity would have to be close to perfect to allow such acceleration. How to change this concept, via the assumed parameters, to bring down the acceleration, is left as an exercise for the reader.

Cost is in the 10 T$ range, even with economies of scale, and is larger than any past human project.

## 4.3 Starship at 0.5c—Getting There Fast

Pushing into the relativistic realm, with a gamma factor of 1.15, shows the enormous energies required: kinetic energy is 4.5 x 10$^{21}$ J. Millis shows that such propulsion energies may not be available for centuries [23]. Because the sail is relativistic, the force on it will be reduced. McInnes describes the relativistic relations [2]. The correction is roughly $\gamma/1-\beta$, about a factor of two. We increase the sail to 100 km diameter; consequently mass becomes 400 tons. Using a laser, with perfect reflectivity and present-day cost parameters (Table 3), the acceleration is low, and occurs over half a light year. Consequently, pointing accuracy is very high, much less than a nanoradian. Capital cost, while high, is only 2.7 times that of the 0.1 c example. Costs of electricity to drive the laser, albeit at today's rates, now greatly exceed capital cost. In fact, the laser is only 0.5% of the cost, so the choice of frequency doesn't matter.

| Starship Concept | Interstellar Precursor | 0.1 c Starship | 0.5 c Starship |
|---|---|---|---|
| *Assumed Parameters* | | | |
| velocity | $6.3 \times 10^4$ m/s (63 km/sec) | $3 \times 10^7$ m/s | $1.5 \times 10^8$ m/s |
| mass | 63 kg | 4,000 kg (4 tons) | $4 \times 10^5$ kg (400 tons) |
| Sail diameter $D_s$ | 1 km | 10 km | 100 km |
| frequency | 100 GHz | 100 GHz | $3 \times 10^{14}$ ($\lambda$ =1 µm) |
| *Calculated Physical Parameters* | | | |
| kinetic energy | $1.25 \times 10^{11}$ J | $1.8 \times 10^{18}$ J | $4.5 \times 10^{21}$ J |
| power | 38 GW | 486 TW | 120 TW |
| antenna aperture | 11.4 km$^2$ | $9 \cdot 10^3$ km$^2$ | $1.67 \cdot 10^4$ km$^2$ |
| antenna diameter | 3.8 km | 430 km | 146 km |
| sail acceleration | 2.4 m s$^2$ | 770 m/s$^2$ | 1.9 m/s$^2$ |
| acceleration distance | $5.2 \cdot 10^8$ m=520,000 km | $6 \cdot 10^{11}$ m=3.9AU | $6 \cdot 10^{15}$ m= 0.6 ly |
| acceleration time | 4.6 hours | 11 hours | 920 days |
| pointing accuracy | 2 µ rad | 67 n rad | 0.02 n rad |
| Efficiency (v/c) | 0.02% | 10% | 50% |
| *Calculated Optimized Costs w/o economies of scale* | | | |
| Total Capital Cost, $C_C$ | 230 B$ | 2900 T$ | 33.500 T$ |
| Operating Cost, $C_O$ | 41 M$ | 1.3 T$ | 2,500 T$ |

Table 1 Starship concepts parameters. Efficiency is that of conversion of electricity to sail kinetic energy. Costs do not include economies of scale, which are given in Table 2.

| *Economies of Scale Costs* | **Interstellar Precursor** | **0.1 c Starship** | **0.5 c Starship** |
|---|---|---|---|
| Power cost | 10 B$ | 14 T$ | 44 T$ |
| Antenna cost | 20 B$ | 28 T$ | 70 T$ |
| Total Capital Cost | 30 B$ | 42 T$ | 114 T$ |
| Cost Savings Ratio | 7.7 | 70 | 295 |

Table 2 Starship concepts costs with economies of scale taken into account (Eq.19). Cost parameters for power and area based on coefficients in Table 3. Operating cost does not change with economies of scale, see Eq. 18.

## 5. Observations on Cost-Optimized Scaling

1) <u>Optimum scaling</u> The cost elements, antenna and power source, are both proportional to the same key features, the velocity, transmitter diameter and frequency:

$$C_C^{opt} \propto \left[\frac{mV_0^2}{D_s f}\right]^{2/3} \tag{20}$$

Therefore, cost can be reduced by
- larger sails,
- lower mass sails,
- higher frequency, probably the upper mm wave >100 GHz.

2) <u>Most important cost</u> Note that p, the cost per watt, is more important than a, the cost per square meter.

$$C^{opt} \propto p^{2/3} a^{1/3} \tag{21}$$

So, reducing the cost of power will be more important than reducing the cost of antennas.

3) <u>Transit time scaling</u> Flight time to the target star $\tau$ is

$$\tau \propto \frac{1}{V_0} \propto \frac{1}{C_{opt}^{3/4}}$$

$$C_{opt} \propto \frac{1}{\tau^{4/3}} \tag{22}$$

So, halving the transit time by doubling the speed will cost 2.5 times as much. That's an unfavorable scaling.

4) One clear conclusion from these examples is that the high speeds necessary for fast interstellar missions make the operating cost exceed the capital cost. Missions will not be flown until the cost of electrical power in space is reduced orders of magnitude below current levels.

5) <u>Power scales like area</u>  Note the linear proportionality between the optimal power and optimal area. To maintain minimum cost while increasing the effective isotropic radiated power, both must be increased in proportion.

6) <u>Large sail scaling</u> In present-day small sails, mass is mostly payload. But much larger sails will have most mass in the sail material. For that case, the technology parameter that drives performance for sails is the area mass density σ, in kg/m², m= σ $A_s$ . Rewriting Eq. 12,

$$C_C^{opt} \propto \left[\frac{\sigma D_s V_0^2}{f}\right]^{2/3} \propto D_s^{2/3} \qquad (23)$$

We should move toward smaller sails with as low an area mass density as possible.

7) <u>Cost vs speed</u> For faster interstellar precursor probes, Eq. 12 means that cost will scale as $v^{4/3}$.

8) <u>Trading off capital vs. operating costs</u>:  The ratio of capital cost to operating cost is

$$\frac{C_C^{opt}}{C_o} \propto \frac{1}{mv}\left[\frac{mv^2}{D_s}\right]^{2/3} = \left[\frac{v}{mD_s^2}\right]^{1/3} \qquad (24)$$

Note this *does not include economies of scale*. If the sail dominates the mass, as would likely occur for more advanced missions, m= σ π$D_s^2$/4, and the scaling becomes

$$\frac{C_C^{opt}}{C_o} \propto \left[\frac{v}{m^2}\right]^{1/3} \qquad (25)$$

So, the designer can shift the cost to the operating budget by decreasing speed, or increasing mass, either by increasing payload mass or, and most likely, sail diameter.

## 5.1 Frequency Figure of Merit

What is the 'right' frequency for directed energy? Advocates of microwaves point to that fully developed, low-cost technology that propagates easily in atmosphere and vacuum. Millimeter-wave proponents favor present-day high average power capability and better focusing due to higher frequency, although it propagates with little loss only in certain 'windows' in the atmosphere. Laser fans feel that technology will win out because of the much higher frequency leading to much smaller optical elements, though propagation is poor in atmosphere at high intensity.

Here we propose a Figure of Merit (FOM) for directed energy based on cost. The frequency-dependent factors in the optimized capital cost are, from Eq. 12:

$$C^{opt} \propto \left[\frac{p^2 a}{f^2}\right]^{1/3} \equiv \text{FOM} \qquad (26)$$

Table 3 shows three technologies, and the Figure of Merit. The laser technology is a fiber laser with 20% electrical efficiency. Gyrotrons have 50% efficiency with direct converters, magnetrons can be 90% efficient.

| Technology | Power cost p | Aperture Cost a | Frequency | Figure of Merit |
|---|---|---|---|---|
| **Magnetron** | 1 \$/W | 1000 \$/m$^2$ | 1-10 GHz | 2 |
| **Gyrotron** | 3 \$/W | 10,000 \$/m$^2$ | 100 GHz | 2 |
| **Laser** | 140 \$/W | 1M\$/m$^2$ | 3 10$^{14}$ Hz (1 μm) | 1.25 |

Table 3 Capital cost Comparisons of Technologies. Figure of Merit is in units of 10$^{-6}$ \$-sec/W-m$^2$. Millimeter and laser data courtesy of Kevin Parkin and Creon Levit. Costs are based on a 1 MW magnetron, 1 M\$ unit, millimeter wave 1.2 MW unit at 3.6 M\$, and a 1 kW, 1 micron fiber laser bought for 140 k\$. Antenna costs basis are satellite dishes for microwave, 64 ALMA 12-m diameter dishes, and 1-m optic at 1 M\$.

Remarkably, the technologies have equal capital cost figures of merit at today's costs. The focusing advantage of lasers is cancelled by the high cost of both the laser and the aperture (sometimes called beam director', 'telescope'). But note the high electrical costs of driving lasers, due to their low efficiency (Table 1) also makes them have a much larger total (capital plus operating) cost. Consequently, the total cost of millimeter wave systems is lower.

Microwave technology is well developed, and has good costing data. Millimeter wave is a younger technology than lasers; the millimeter market is just beginning to develop so the costs are evolving. High power continuous lasers, after developing quite slowly, are beginning to see prices drop, but still do not have a clear market. They may soon reach a firm price point. At present, the cheaper microwave and millimeter wave are readily available. They are much easier to use for experiments, so those experiments are more likely to be done. Consequently, they are likely to become more practical.

Of course, future cost changes will determine the most cost-effective technology.

**6.0 Development Path for Directed Energy Propulsion**

To eventually have a directed energy capability, space infrastructure must exist to build on. How do we get from where we are now to a future when directed energy can be used for fast missions, including interstellar? By developing solar sails and other applications of directed energy in parallel.

**6.1 Solar Sail Development**

Recent papers by Friedman and Matloff in this Symposium have shown a path for solar sail technology to lead to speeds on the scale of the Interstellar Precursor described in 4.2 [23-25]. Such development directly enables directed energy sails:

- Sail engineering, especially materials: carbon nanotubes,, carbon microtruss
- As sails are large-scale structures in space, they also influence the development of large transmitting antennas.

This leads to:
- Larger sails,
- Lighter sails,
- Faster sails
- Fast Solar Sail Missions, for example to the Oort Cloud, Heliopause, and Interstellar Medium

This prepares the ground for Beam Propulsion.

**6. 2 Development of Directed Energy Propulsion**

Power beaming becomes economic only when it can move power from where it is cheap and accessible to places where it is hard to come by. Previous work has shown that it is often more economical to transmit power than to move the equipment to produce power locally. Modern power systems are complex, but if power for space can be located where it is easily accessible and adjacent to where the required skilled people are located, i.e., on Earth, then it becomes more practical.

Applications can be met by building up a system existing technologies: microwave and millimeter wave antennas are already in use for astronomy, gyrotron sources at high frequencies (>100 GHz) are being developed for fusion. The method is to build, stairstep-like, a sequence of applications of beaming power [18]:

- Orbital debris mapping could be the first objective.

- Recharging of satellite batteries in LEO could be economic [26], followed by recharging of satellites in GEO.

- Launch into orbit of 1000 kg cargo-carrying supply modules makes industrial transport in and out of LEO a reality at cost about an order of magnitude less than present day [27].

In today's frugal climate, it is important for technology development to be coupled to commercial applications. Several of the missions we've described are potentially commercial matters. Starting with orbital debris mapping, one can see an incremental commercial development leading first to satellite power recharging. Eventually, as the space market and business confidence grows and capital becomes more available, this development plan leads to the repowering of satellites in GEO and ultimately to launch services. Investment costs are minimized because the research program leads to many applications.

Therefore, the private sector should be included from the outset in the development of power beaming for space applications. This includes the R&D phase, as it is very important to gain support from industry to maintain a long-term commercial strategy.

There is at present no clear view of how it is to be achieved and by what technology we are to make the Solar System readily accessible. This paper has attempted to demonstrate that the technical means are already in hand for a space infrastructure. A unified approach to many missions can be found by looking to the use of microwave, millimeter wave and laser beams to provide power and transportation. Interplanetary infrastructure development will be treated further in the forthcoming book *Starship Century* [28].

## 7. Conclusions

Microwave propelled sails have *no physics issues* and offer much lower cost probes. Its large-scale antenna and powerful radiator mean the questions to face are engineering and cost. Relations have been derived here for quantifying this question, including economies of scale ('learning curve'). One clear conclusion from examples shown here is that the cost of interstellar sailships is very high, driven by current costs for electrical power, radiation source and antennas. The high speeds necessary for fast interstellar missions make the operating cost exceed the capital cost. *Missions will not be flown until the cost of electrical power in space is reduced orders of magnitude below current levels.*

The usefulness of the beamed power/sail method awaits further quantification by:

- Analyzing past concepts (Forward, Landis, Frisbee, Matloff) to see if they are off-optimal, so can be improved.
- Searching for the lowest cost missions by exploring parameter variations of the physical parameters.

- Quantifying an alternate use of sails-deceleration of sail probes from a fusion-powered starship as it approaches stellar systems.

**8. Acknowledgments** I gratefully acknowledge data from Kevin Parkin and Creon Levit and discussions with Greg Matloff, Greg Benford and Geoff Landis.

**9. References**

1. D. Price, J. Levine, and J. Benford, "ORION-A Frequency-Agile HPM Field Test System", Seventh National Conference on High Power Microwave Technology, Laurel, MD, 1997.

2. C. McInnes,, *Solar Sailing: Technology, Dynamics, and Mission Applications*, Springer-Verlag, NY, 1999.

3. R. L. Forward, "Roundtrip Interstellar Travel Using Laser-Pushed Lightsails", *J. Spacecraft and Rockets*, **21**, pp. 187-195, 1984.

4. R.L. Forward, "Starwisp: an ultra-light interstellar probe", *J. Spacecraft and Rockets*, **22**, pp. 345-350, 1985.

5. J. Benford and G. Benford, "Flight Of Microwave-Driven Sails: Experiments And Applications", *Beamed Energy Propulsion*, AIP Conf. Proc. **664**, pp. 303-312, A. Pakhomov, ed., 2003.

6. E Schamiloglu, "3-D Simulations Of Rigid Microwave Propelled Sails Including Spin", *Proc. Space Technology and Applications International Forum*, AIP Conf. Proc. **552**, 559, 2001.

7. G. Benford, O. Goronostavea and J. Benford, "Experimental Tests Of Beam-Riding Sail Dynamics", *Beamed Energy Propulsion*, AIP Conf. Proc. **664**, pp. 325-335, Pakhomov A., ed., 2003.

8. G. Benford and P. Nissenson, "Reducing Solar Sail Escape Times From Earth Orbit Using Beamed Energy", *JBIS*, **59**, pp. 108-111, 2006.

9. T. R. Meyer, C. P. McKay, P.M. McKenna and W.R. Pyror, "Rapid Delivery of Small Payloads to Mars", AAS Paper AAS-84-172, *The Case for Mars II*, pp 419-431, C.P. McKay, Ed., Vol. **62** of the Science and Technology Series, Am. Astronomical Society, 1985.

10. J. Kare, "Beamed Power Missions to Jupiter", Presented at the Space Exploration and Development Horizon Mission Methodology Workshop, Jet Propulsion Laboratory, Pasadena CA, September 22-23, 1994.

11. G. Benford and J. Benford, "Power-Beaming Concepts For Future Deep Space Exploration", *JBIS*, **59**, pp. 104-107, 2006.

12. G.A Landis, "Small Laser-Propelled Interstellar Probe", *JBIS*, **50**, pp 149-154, 1977.

13. G.A Landis, "Optics and Materials Considerations for a Laser-Propelled Lightsail", IAA Paper IAA-89-684, Presented at the 40th Congress of the Intl. Astronautical Federation, 1989.

14. G.A Landis, "Beamed Energy Propulsion For Practical Interstellar Flight", *JBIS*, **52**, 420, 1999.

15. G.A Landis, "Microwave-Pushed Interstellar Sail: *Starwisp* Revisited", paper AIAA-2000-3337, 36th Joint Propulsion Conference, 2000.

16. R.H. Frisbee, "Limits of Interstellar Flight Technology", in *Frontiers of Propulsion Science*, **Vol. 227**, ed. M. G. Millis and E. W. Davis, Progress in Astronautics and Aeronautics, AIAA Press, Reston VA, p.31, 2009.

17. J. Benford, D. Benford and G. Benford, "Messaging With Cost Optimized Interstellar Beacons", *Astrobiology*, **10**, pp. 475-491, 2010. Also at arxiv.org/abs/0810.3964v2

18. J. Benford and R. Dickinson, *S*pace propulsion and power beaming using millimeter systems*, Proc. SPIE*, 2557, pp. 179-193, 1995. Also published in *Space Energy and Transportation*, **1**, 211, 1996.

19. J. Kare, and K. Parkin, "Comparison of laser and microwave approaches to CW beamed energy launch" In *Beamed Energy Propulsion-2005*, K. Komurasaki, ed., AIP Conf. Proc. **830**, New York: Am. Inst. of Physics, pp. 388-399, 2006.

20. N. Luhmann, et al., "Affordable Manufacturing", *Modern Microwave and Millimeter-Wave Power Electronics*, Ch. 14, ed. R. Barker et al, IEEE Press, Piscataway, N.J., pp.731-764, 2005.

21. P.C. Liewer, R.A. Mewaldt, J.A. Ayon, C. Gamer, S. Gavit, R.A. Wallace, "Interstellar probe using a solar sail: conceptual design and technological challenges", in: K. Scherer, H. Fichtner, H.-J. Fahr, E. Marsch (Eds.), COSPAR Colloquium on The Outer Heliosphere: The Next Frontiers COSPAR Colloquia Series, Pergamon Press, New York, pp. 411–420, 2001.

22. G.L. Matloff, "The Beryllium Hollow-Body Sail and Interstellar Travel", *JBIS*, **59**, pp. 349-354 2006.

23. M. Millis, "First Interstellar Missions, Considering Energy and Incessant Obsolescence", JBIS, **63**, pp. 434-443, 2010.

24. L. Friedman, D. Garber, and T. Heinsheime, "Evolutionary Lightsailing Missions for the 100-Year Starship", *JBIS*. Submission pending, 2011.

25. G. L. Matloff, "Interstellar Light Sails", *JBIS*. Submission pending, 2011.

26. Y. K. Bae, 'Photonic Railway: "A Sustainable Developmental Pathway Of Photon Propulsion", *JBIS*. Submission pending, 2011.

26. M., Miller G., Kadiramangalam M. and W. Ziegler, "Earth-to-satellite microwave power transmission", *J. of Propulsion & Power*, **5**, pp. 750, 1989.

27. K. Parkin, L.D. DiDomenico and F.E.C. Culick, The microwave thermal thruster concept, in *AIP Conf. Proc.702 Second international symposium on beamed-energy propulsion*, Komurasaki K., Ed., Melville NY, 418, 2004.

28. G. Benford and J. Benford, *Starship Century*, in press, 2013.

**Appendix A**

**Cost Optimization**

Examples of how cost optimization can make a big difference:

1) <u>Interstellar Beacons</u> We used this method to quantify the cost of interstellar beacons, the subject of SETI. The result shows that there is a sharp cost optimum, as in Figure A1.

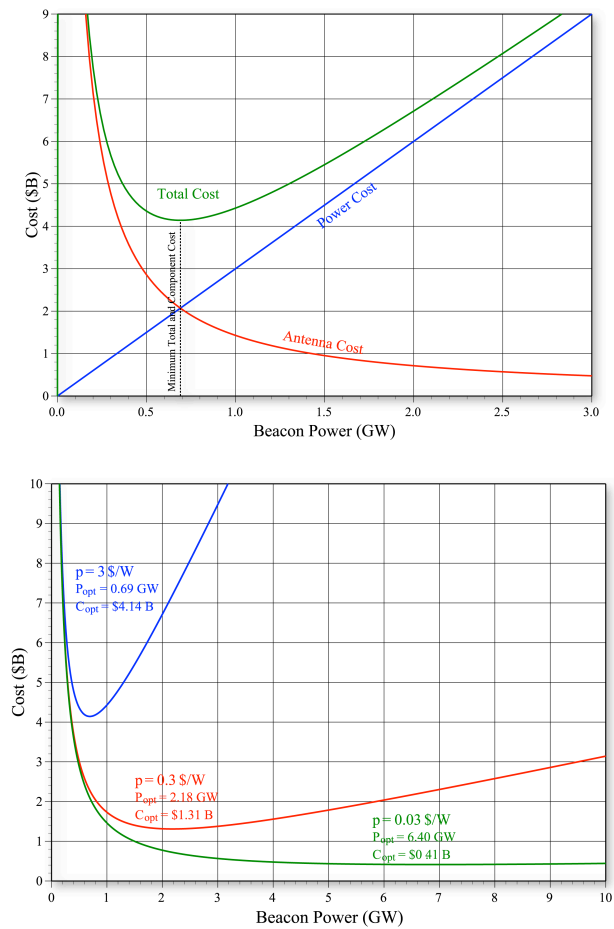

**Fig. A1a.** Antenna, microwave power and total costs of Beacon of EIRP=$10^{17}$ W, antenna cost k$/m$^2$, power cost p=3 $/W, at f=1 GHz. Note: minimum total cost is at the point where the antenna cost and power cost are equal. **Fig. A1b.** Impact of pulsed sources on system cost of Beacons of EIRP=$10^{17}$ W. Cases of long-pulse sources at 3$/W (same as Fig. A1a case) and short-pulse sources at 0.3$/W and 0.03$/W. [17]

2) Beamed Energy Launcher  It is possible to estimate the size of a millimeter-wave beam-driven rocket, which minimizes beam facility cost. This minimum occurs at a balance between two opposing expenses: the antenna/optics and the millimeter wave power (Fig. 2).

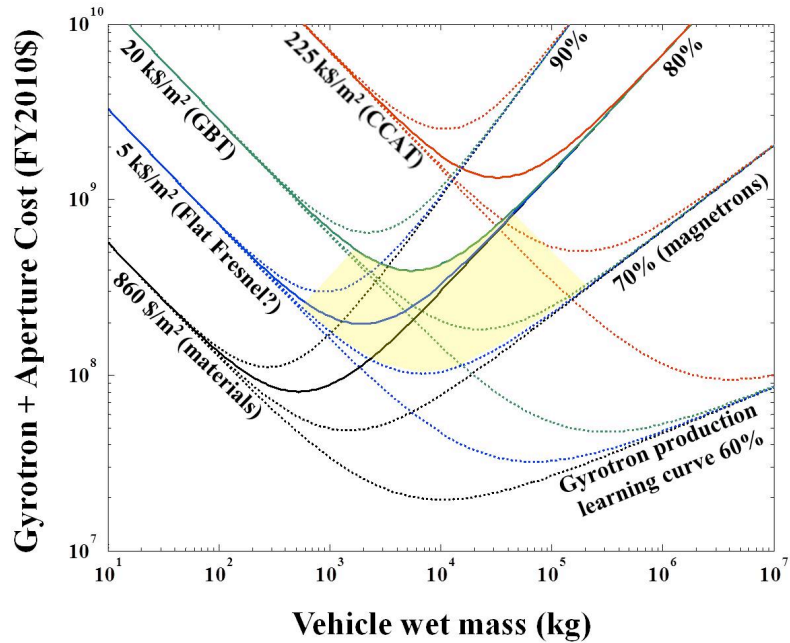

**Fig. A2**. Tradeoff in minimizing cost of millimeter wave beam-driven rocket shows that cost-optimization makes a big difference (K. Parkin, private communication). Note that cost optimized (minimized) cost implies minimized mass as well, as antenna area and beamed power are proportional to both cost and mass.

**Appendix B**

**Sail Starship Equations of Motion**

For a beam-driven sail, at what range does the beamwidth size exceed the sail size? If power is constant in time, what speed is attained at this point? How much more speed results if the beam remains on beyond this point?

When the diameter of the sail $D_s$ is equal to the spot size of the beam at $R_0$,

$$2.44 \lambda / D_t = D_s/R_0, \qquad (A1)$$

$$R_0 = D_s D_t / 2.44\lambda$$

Note that at this point power transfer can be quite efficient. Force on the sail will be constant out to $R_0$, and then will fall off as $R^{-2}$. Denote the force for $R < R_0$ as $F_0 = [\eta +1] P/c$, where P is the power, $\eta$ the reflectivity. To find the speed at any range, solve the equation of motion and solev for the constant force region and that where force varies as 1/R:

$$a = \frac{dv}{dt} = \frac{dv}{dr}\frac{dr}{dt} = v\frac{dv}{dr} = \frac{F}{m}$$

$$\int_0^v v\, dv = \int_0^R \frac{F(R)}{m} dr \qquad (A2)$$

$$F = F_0, R < R_0$$

$$F = F_0 \left(\frac{R_0}{R}\right)^{2,}, R > R_0 \qquad (A3)$$

For $R = R_0$,

$$\frac{v_0^2}{2} = \frac{F_0 R_0}{m}$$

$$v_0 = \sqrt{\frac{2 F_0 R_0}{m}} = \sqrt{\frac{2(\eta+1) P R_0}{mc}} \qquad (A4))$$

For $R > R_0$,

$$\int_{v_0}^{v_f} v\, dv = \int_{R_0}^{\infty} \frac{F_0}{m}\left(\frac{R_0}{R}\right)^2 dR$$

$$\frac{v_f^2}{2} - \frac{v_0^2}{2} = \left(\frac{F_0}{m} R_0^2\right)\left[\frac{1}{\infty} - \frac{1}{R_0}\right] = \frac{F_0}{m} R_0 \qquad (A4)$$

$$v_f = 2\sqrt{\frac{F_0 R_0}{m}} = 2\sqrt{\frac{(\eta+1) P R_0}{mc}} = \sqrt{2} v_0$$

Continuing to drive the sailship beyond $R_0$ makes sail velocity increase by $2^{1/2}-1=41\%$. The energy can be doubled if the sail is accelerated far beyond $R_0$. But the efficiency gradually falls as the beam gets ever larger than the sail.

**Appendix C**

**Sail Starship Design Methodology.**

The concepts in Table 1 are derived from the optimization relation (Eq. 12) and kinematics. The procedure is:

Assume Key Parameters

1) Assume velocity v and mass m.

2) Assume sail diameter $D_s$. This is important because 1) It should fit with the area mass density $\sigma$, m=$\sigma$ A. See fifth remark in Section 5. 2) From Eqs. 3 and 12, acceleration a~$1/D^2$. A small sail can give accelerations sufficient to melt or sublime the sail material, so estimate the limiting acceleration and stay below it. As this involves Eq. 14, you may have to iterate.

3) Assume a frequency domain. Note that assigns the cost parameters for power and area, p and a (Eq. 5, Table 2).

Calculate Physical Parameters

4) Calculate kinetic energy KE (=$mv^2/2$)

5) Calculate optimized cost (economies of scale costs are calculated later) from Eq. 12.

6) Calculate optimum area (and diameter) and power from Eqs. 13 and 14.

7) Calculate acceleration from Eq. 3.

8) Calculate range $R_0$ from Eq. A1.

9) Calculate acceleration time (beam-on time) from

$$t = \sqrt{\frac{2R_0}{a}} \qquad (A5)$$

Calculate Costs

10) Calculate operating cost from Eq. 18.

11) Calculate economies of scale from Eq. 19. The key parameter is N, which is either the number of sources ($N_s$) or the number of antenna (optic) elements ($N_a$). Use the data from Table 2.

12) Calculate a reduced capital cost from economies of scale by adding results of 10) and 11).